\documentclass[
aps,%
12pt,%
final,%
notitlepage,%
oneside,%
onecolumn,%
nobibnotes,%
nofootinbib,% 
superscriptaddress,%
noshowpacs]%
{revtex4}%
\usepackage{graphics}
\usepackage[english]{babel}
\usepackage{url}
\usepackage{graphicx}
\usepackage{xcolor}
\usepackage{amsmath}
\usepackage{amssymb}
\usepackage{slashed}
\usepackage{multirow}
\usepackage{caption}
\usepackage{cleveref}
\allowdisplaybreaks

\begin{document}

\title{Production of $D$-wave states of $\bar b c$ quarkonium  at the LHC}

\author{\firstname{A.~V.}~\surname{Berezhnoy}}
\email{Alexander.Berezhnoy@cern.ch}
\affiliation{SINP MSU, Moscow, Russia}

\author{\firstname{I.~N.}~\surname{Belov}}
\email{ilia.belov@cern.ch}
\affiliation{SINP MSU, Moscow, Russia}
\affiliation{Physics department of MSU, Moscow, Russia}

\author{\firstname{A.~K.}~\surname{Likhoded}}
\email{Anatolii.Likhoded@ihep.ru}
\affiliation{NRC Kurchatov Institute IHEP, Protvino, Russia}

\begin{abstract}
The hadronic production of $D$-wave states of $\bar b c$ quarkonium is studied. The relative yield of such states is estimated for kinematic conditions of LHC experiments.
\end{abstract}

\maketitle
\section{Introduction}
\label{sec:intro}

Recently $2S$ excitations of  $B_c$ mesons have been discovered  by CMS~\cite{Sirunyan:2019osb, Sirunyan:2020mzn,CMS:2020wnn} in $B_c \pi^+\pi^-$ spectrum. The observation has been confirmed by LHCb experiment~\cite{Aaij:2019ldo}. Thus, in the opinion of the authors of~\cite{Eichten:2019gig}, it has opened a new era in the spectroscopy of ordinary quarkonia. The excellent experimental results, the long known theoretical prediction, that $D$ states  can also  decay to $B_c \pi^+\pi^-$  with probability $\sim 20$\%   \cite{Eichten:1994gt} (see also \cite{Eichten:2019gig}), and the earlier  study of $D$-wave quarkonia production within fragmentation approximation~\cite{Cheung:1995ir} stimulate us to estimate the cross section of $D$-wave $B_c$ states in hadronic interactions.  It is worth to note, that the relative $B_c(2S)$ yield $\sigma(B_c(2S)/\sigma(B_c)$  published by CMS ($\sim 8 \%$) is in a good agreement with our prediction ($\sim 10 \%$)~\cite{Berezhnoy:2013sla,Berezhnoy:2019yei}. That is why we hope, that our prediction for relative yield of $D$-wave states obtained within analogous technique will fairly good describe the future experimental observation of the discussed states. 

The article is organized as follows: Section II is devoted to the calculation technique description; in Sec. III the relative yield of $D$-wave $B_c$ meson state is estimated for kinematic conditions of the LHC experiments; in  Sec. IV we make  conclusions on the possibility observation of such states at LHC; in the Appendix we provide some information about $D$-wave $B_c$ masses and wave functions.

\section{Calculation technique}

To estimate the production amplitude of $D$-wave $B_c$ states we use the analogous technique as for $S$ and $P$ waves,  namely, we perform calculations  within the  color singlet model neglecting the internal velocities of quarks inside quarkonium (see for details \cite{Berezhnoy:1994ba,Chang:1994aw, Berezhnoy:1995au,Kolodziej:1995nv,Berezhnoy:1996ks,Berezhnoy:1997fp,Baranov:1997sg, Baranov:1997wv, Berezhnoy:1997uz, Chang:2004bh, Berezhnoy:2004gc,Chang:2005wd,Chang:2006xka,Berezhnoy:2010wa,Gao:2010zzc}): 
\begin{equation}
%\notag
A \sim 
\int d^3q\,\Psi^*({\boldsymbol q})\left\{ \bigl.T(p_i,{\boldsymbol q}) \bigr|_{\boldsymbol q=0}+
\bigl.{q^\alpha}\frac{\partial}{\partial {q^\alpha}} T(p_i,\boldsymbol q) \bigr|_{\boldsymbol q =0} + {\frac{1}{2} \bigl.{q^\alpha q^\beta}\frac{\partial^2}{\partial {q^\alpha} \partial {q^\beta}} T(p_i,\boldsymbol q) \bigr|_{\boldsymbol q =0}} +  \dotsb \right\},
\label{eq:amp_general}
\end{equation}
where $T$ is the amplitude of four heavy quark gluonic production with momenta   $p_i$ in the leading-order approximation, which is contributed by 36 Feynman diagrams;  ${\boldsymbol q}$  is the quark three-momentum in the $B_c$ meson rest frame,  and  $\Psi({\boldsymbol q})$ is the $B_c$ meson wave function.

For  $D$-wave states the first two terms in \eqref{eq:amp_general} are equal to zero, and therefore an amplitude is proportional to the second derivative of the wave function at origin $R''(0)$ and to the second derivatives of  $T$ over  ${\boldsymbol q}$. The amplitudes for the spin singlet  $A^{j_z}$ $(J=2,\ j_z=l_z)$ and for the spin triplet $A^{Jj_z}$  $(J=1,2,3;\ j_z = s_z+l_z)$  can be 
expressed as follows (see also~\cite{Cheung:1995ir}, where the $D$-wave $B_c$  production was studied in the fragmentation approach):
\begin{equation}
A^{j_z} = \frac{1}{2}\sqrt{\frac{15}{8\pi}}R_D''(0)\epsilon^{\alpha\beta}(j_z)\left.\frac{\partial^2 M(\boldsymbol{q})}{\partial q^{\alpha} \partial q^{\beta}}\right|_{\boldsymbol{q}=0},
\end{equation}
\begin{equation}
A^{Jj_z} = \frac{1}{2}\sqrt{\frac{15}{8\pi}}R_D''(0)\Pi^{J,\ \alpha\beta\rho}(j_z)\left.\frac{\partial^2 M_{\rho}(\boldsymbol{q})}{\partial q^{\alpha} \partial q^{\beta}}\right|_{\boldsymbol{q}=0} ,
\end{equation}
where  
\begin{equation}
\Pi^{J,\ \alpha\beta\rho}(j_z) = \sum_{l_z, s_z}\epsilon^{\alpha\beta}(l_z)\epsilon^\rho(s_z)\cdot C^{Jj_z}_{s_zl_z},
\end{equation}
 $\epsilon^{\rho}$ and $\epsilon^{\alpha\beta}$ are vector and polarization tensors and  $C^{J j_z}_{s_z l_z}$ are  Clebsch-Gordan coefficients.

The states with a definite spin value  are constructed by operators 
\begin{equation}
{\cal P}(0,0) = \frac{1}{\sqrt 2}\lbrace v_+(p_{\bar b} + k)\overline{u}_+(p_c - k) - v_-(p_{\bar b} + k)\overline{u}_-(p_c - k)\rbrace 
\label{eq:projector_0}
\end{equation}
and
\begin{equation}
{\cal P}(1,s_z) = \begin{cases}
{\cal P}(1,1) =  v_-(p_{\bar b} + k)\overline{u}_+(p_c - k) \\
{\cal P}(1,0) =  \frac{1}{\sqrt 2}\lbrace v_+(p_{\bar b} + k)\overline{u}_+(p_c - k) + v_-(p_{\bar b} + k)\overline{u}_-(p_c - k)\rbrace \\
{\cal P}(1,-1) = v_+(p_{\bar b} + k)\overline{u}_-(p_c - k) \\
\end{cases}.
\label{eq:projector_1}
\end{equation}

The spinors in \eqref{eq:projector_0} and\eqref{eq:projector_1} are expressed as follows:
\begin{align}
&v_{\lambda_1}(p_{\bar b} + k) = (1 - \frac{\slashed k}{2m_b})v_{\lambda_1}(p_{\bar b}), \\
&\overline{u}_{\lambda_2}(p_c - k) = (1 - \frac{\slashed k}{2m_c})\overline{u}_{\lambda_2}(p_c),
\end{align}

where $p_{\bar b}=\frac{m_b}{m_b+m_c}P_{B_c}$, $p_c=\frac{m_c}{m_b+m_c}P_{B_c}$ and $k(\boldsymbol q)$ is a Lorentz boost of four-vector $(0, \boldsymbol q)$ to the system where the $B_c$ momentum is equal to    $P_{B_c}$.

\vspace*{0.3cm}
Amplitudes  and  their derivatives have been calculated numerically.
%as below:
%
%\begin{equation}
%\frac{\partial^2 M}{\partial q_x^2 }\approx \frac{M(p_i, q_x)+M(p_i, - q_x)-2 M(p_i, 0)}{\Delta^2},
%\end{equation}
%\begin{equation}
%\frac{\partial^2 M}{\partial q_x \partial q_y }\approx %\frac{M(p_i, q_x+ q_y)+M(p_i, 0)- M(p_i, q_x)-M(p_i,  %q_y)}{\Delta^2}
%\end{equation}
To simplify the calculations we square and summarize amplitudes, keeping only a spin value $S=0$ or $S=1$. The amplitude squared for the spin-singlet state with $S=0$ ($1 ^1D_2$)  is given by the following equation:
\begin{multline}
 %|A_{S=0}|^2 
\mathbb{P}= \left(\frac{5}{16\pi}\right)|R_D^{''}(0)|^2\times\\  
\left[
\left( 
\left|\frac{\partial^2 M_{S=0}}{\partial q_x^2}\right|^2 +
\left|\frac{\partial^2 M_{S=0}}{\partial q_y^2}\right|^2 + \left|\frac{\partial^2 M_{S=0}}{\partial q_z^2}\right|^2 \right)\right.\\
\left.
 + 3 
\left(
\left|\frac{\partial^2 M_{S=0}}{\partial q_x \partial q_y}\right|^2 + 
\left|\frac{\partial^2 M_{S=0}}{\partial q_x \partial q_z}\right|^2 + 
\left|\frac{\partial^2 M_{S=0}}{\partial q_y \partial q_z}\right|^2 
\right)
\right.
\\ 
 - \left.
 \text{Re}\left(\frac{\partial^2 M_{S=0}}{\partial q_x^2}\frac{\partial^2 M^*_{S=0}}{\partial q_y^2} + \frac{\partial^2 M_{S=0}}{\partial q_x^2}\frac{\partial^2 M^*_{S=0}}{\partial q_z^2} +
\right.\right.
\\
+ \left.\left. \frac{\partial^2 M_{S=0}}{\partial q_y^2}\frac{\partial^2 M^*_{S=0}}{\partial q_z^2}\right)\right]. 
\end{multline}

The sum of amplitudes squared for the spin-triplet states with $S=1$  ($1 ^3D_1$, $1 ^3D_2$, $1 ^3D_3$) is presented  below:
\begin{multline}
% \sum_{j}^{-1,0,1}|A_{S=1,s=j}|^2 =
\mathbb{V}=
\left(\frac{5}{16\pi}\right)|R_D^{''}(0)|^2\times\\  
\sum_{s_z}^{-1,0,1}\left[
\left( 
\left|\frac{\partial^2 M_{S=1,s_z}}{\partial q_x^2}\right|^2 +
\left|\frac{\partial^2 M_{S=1,s_z}}{\partial q_y^2}\right|^2 + \left|\frac{\partial^2 M_{S=1,s_z}}{\partial q_z^2}\right|^2 \right)\right.\\
\left.
 + 3 
\left(
\left|\frac{\partial^2 M_{S=1,s_z}}{\partial q_x \partial q_y}\right|^2 + 
\left|\frac{\partial^2 M_{S=1,s_z}}{\partial q_x \partial q_z}\right|^2 + 
\left|\frac{\partial^2 M_{S=1,s_z}}{\partial q_y \partial q_z}\right|^2 
\right)
\right.
\\ 
 - \left.
 \text{Re}\left(\frac{\partial^2 M_{S=1,s_z}}{\partial q_x^2}\frac{\partial^2 M^*_{S=1,s_z}}{\partial q_y^2} + \frac{\partial^2 M_{S=1,s_z}}{\partial q_x^2}\frac{\partial^2 M^*_{S=1,s_z}}{\partial q_z^2} +
\right.\right.
\\
+ \left.\left. \frac{\partial^2 M_{S=1,s_z}}{\partial q_y^2}\frac{\partial^2 M^*_{S=1,s_z}}{\partial q_z^2}\right)\right].
\end{multline}

A more rigorous consideration of this process within NRQCD~\cite{Bodwin:1994jh} implies that the final meson is no longer a $\bar b c$ pair rather a superposition of Fock states:
\begin{align}
|B_c(1^1D_2)\rangle = O(1)|\bar b c( ^1D_2 , \boldsymbol{1})\rangle + O(v) |\bar b c( ^1P_{1}, \boldsymbol{8})g\rangle  + O(v^2 )|\bar b c( ^1S_0 , \boldsymbol{8} \mbox{ or }
\boldsymbol{1})gg \rangle + \cdots \label{eq:Fock_states_S0}
 \\
|B_c(1^3D_j)\rangle = O(1)|\bar b c( ^3D_j , \boldsymbol{1})\rangle + O(v) |\bar b c( ^3P_{j'}, \boldsymbol{8})g\rangle  + O(v^2 )|\bar b c( ^3S_1 , \boldsymbol{8} \mbox{ or }
\boldsymbol{1})gg \rangle + \cdots
\label{eq:Fock_states_S1}
\end{align}
where $\boldsymbol{1}$ and $\boldsymbol{8}$ refer to color singlet and color octet states
of the quark pair.~\footnote{As  noted in~\cite{Cheung:1995ir},  in the above Fock state expansion there are also other $ O(v^2 )$ states, but their production will be further suppressed by powers of $v$.}
The NRQCD local 4-fermion operators  ${\cal O}^{B_c(1^1 D_2)}_{\boldsymbol{1}} (^1 D_2)$ and ${\cal O}^{B_c(1^3 D_j)}_{\boldsymbol{1}} (^3 D_j)$ relevant for the fist terms in the expansions \eqref{eq:Fock_states_S0} and \eqref{eq:Fock_states_S1}  are related to the quarkonium  wave function $R_D^{''}(0)$:~\footnote{There are two widely used normalizations for  $\mathcal{O}_{1}$ matrix elements. One  normalization method (BBL) inherits from the study~\cite{Bodwin:1994jh}. Another one (PCGMM) is based on the  work~\cite{Petrelli:1997ge}. These two normalization methods relate to each other as follows: $\mathcal{O}_{1}^{\textrm{PCGMM}}=\frac{1}{2N_c}\mathcal{O}_{1}^{\textrm{BBL}}$. Since we consider our study to be a continuation of work~\cite{Cheung:1995ir} that uses the BBL normalization, we also use it in  Eqs.~\eqref{eq:O1S0} and \eqref{eq:O1S1}. The exact determination of the discussed operators one can find for example in \cite{Fan:2009cj} and \cite{Sang:2015lra}.}
\begin{align}
\langle{\cal O}^{B_c(1^1 D_2)}_{\boldsymbol{1}} (^1 D_2)\rangle &\approx \frac{75N_c}{4\pi}|R_D^{''}(0)|^2, 
\label{eq:O1S0}\\
\langle{\cal O}^{B_c(1^3 D_j)}_{\boldsymbol{1}} (^3 D_j)\rangle&\approx \frac{15(2j+1)N_c}{4\pi}|R_D^{''}(0)|^2.
\label{eq:O1S1}
\end{align}
Therefore production of the first components in the expansions \eqref{eq:Fock_states_S0} and \eqref{eq:Fock_states_S1} can be described by  formulas obtained from \eqref{eq:amp_general}. 

By analogy with the fragmentation case~\cite{Cheung:1995ir}, it can be shown that within NRQCD the contributions to the gluonic production of the second and third terms in \eqref{eq:Fock_states_S0} and \eqref{eq:Fock_states_S1}  are of the same order on $\alpha_s$ and $v$, as the contribution of the first terms,  and therefore they  should  be also included.  Having an experience in calculation of gluonic production of $S$- and  $P$-wave  $\bar b c$ color singlets, it is not difficult to calculate the hard parts of appropriate production amplitudes for $S$- and  $P$-wave  $\bar b c$ color octets. Unfortunately the soft part of such amplitudes can not be accurately  estimated, because values of the relevant NRQCD operators are unknown. However understanding the kinematic behavior of such contributions even without knowing the exact normalization  could help in the experimental search of $D$-wave $B_c$ states. We estimate here three additional NRQCD contributions enumerated  in Eqs.~\eqref{eq:Fock_states_S0} and \eqref{eq:Fock_states_S1} using the fairly defined hard parts of the processes normalized by coefficients which are extracted using a naive velocity scaling, as explained below.

To model  the contributions of  $P$-wave color octet states $|\bar b c( ^1P_{1}, \boldsymbol{8})g\rangle$ and $|\bar b c( ^3P_{j}, \boldsymbol{8})g\rangle$ to the discussed cross sections we use our tools for calculation of $P$-wave $B_c$ color singlet states. We replace the color singlet  wave function to color octet  one and multiply the wave function derivative squared at origin $|R_P'(0)|^2$ by  the coefficient $K_{P\bf{8}}$ which is of order $O(v^2_{\mbox{eff}})$, where  $v^2_{\mbox{eff}}$ is an  effective squared velocity of quarks in the $B_c$ meson:
\begin{equation}
\begin{array}{ccc}
\displaystyle 
   \frac{\delta_{\bar bc}}{\sqrt{3}}  & \to & \sqrt{2} \ t_{\bar bc}^a, \\
  |R_P'(0)|^2   & \to  &  K_{P\bf{8}}\cdot |R_P'(0)|^2.
\end{array}
\label{eq:nrqcd-P8g}
\end{equation}

To model the contributions of  $S$-wave color octet states $|\bar b c( ^1S_{0}, \boldsymbol{8})gg\rangle$ and $|\bar b c( ^3S_{1}, \boldsymbol{8})gg\rangle$  we use our tools for calculation of $S$-wave $B_c$ color singlet states, where we replace the color singlet wave function to color octet one and multiply the  coordinate wave function squared at origin $|R_S(0)|^2$ by the coefficient  $K_{S\bf{8}}$ which is of order  $O\left(\left[v^2_{\mbox{eff}}\right]^2\right)$:
\begin{equation}
\begin{array}{ccc}
\displaystyle 
   \frac{\delta_{\bar bc}}{\sqrt{3}}  & \to & \sqrt{2} \ t_{\bar bc}^a, \\
  |R_S(0)|^2   & \to & K_{S\bf{8}}\cdot |R_S(0)|^2.
\label{eq:nrqcd-S8gg}
\end{array}
\end{equation}

Constructing the contributions of  $S$-wave color singlet states $|\bar b c( ^1S_{0}, \boldsymbol{1})gg\rangle$ and $|\bar b c( ^3S_{1}, \boldsymbol{1})gg\rangle$, as in the previous case (\ref{eq:nrqcd-S8gg}) we just rescale the wave function squared at origin $|R_S(0)|^2$ by the coefficient $K_{S\bf{1}}$ which is of order  $O\left(\left[v^2_{\mbox{eff}}\right]^2\right)$:
\begin{equation}
\begin{array}{ccc}
\displaystyle 
  |R_S(0)|^2   & \to & K_{S\bf{1}} \cdot |R_S(0)|^2.
\end{array}
\label{eq:nrqcd-S1gg}
\end{equation}

The very similar approach was applied to estimate the color octet contribution to the $B_c$ $P$-wave production in the work~\cite{Chang:2005bf}. Also in~\cite{Chang:2005bf} the properties of color matrix for the gluonic $\bar b c$ color octet production were studied in details.

It is worth to remind that for gluonic $b\bar b c\bar c$  production the replacement of the color singlet wave function of $\bar b c$-pair to the color octet wave function cannot be reduced to a simple scaling of the matrix element, because it essentially changes the relative contributions of different Feynman diagrams to the total amplitude.~\footnote{For the first time the color matrix for the process $gg\to  b \bar b c \bar c$ was investigated in \cite{Barger:1991vn}, where it was shown that such color matrix has 13 nonzero eigenvalues. Three of them correspond to the cases, where  the $\bar b c$-pair  is in a color singlet state: $(\bar b c)_{\bf{1}}\otimes (b \bar c)_{\bf{1}}$,  $(\bar b c)_{\bf{1}}\otimes (b \bar c)_{\bf{8-symmetric}}$ and $(\bar b c)_{\bf{1}}\otimes (b \bar c)_{\bf{8-antisymmetric}}$. The remaining ten eigenvalues correspond to the cases, where the $\bar b c$-pair  is in a color octet. We refer to the studies \cite{Chang:2005bf,Berezhnoy:2004gc} for details.}

The $v^2_{\mbox{eff}}$  value we estimate as: 
\begin{equation}
v^2_{\mbox{eff}}= \frac{\langle E\rangle}{2\mu},
\label{eq:v2_eff}
\end{equation}
where $\langle E\rangle$  is the averaged kinematic energy of quark inside the $B_c$-meson and $\mu=\frac{m_c m_b}{m_c+m_b}$.  Using  the value  $\langle E\rangle \approx 0.35$~GeV  estimated in \cite{Gershtein:1994jw}, we obtain that   $v^2_{\mbox{eff}}\approx 0.15$.

To estimate the additional NRQCD contributions numerically we choose the following central values for $K$ coefficients in Eqs.~\eqref{eq:nrqcd-P8g} to \eqref{eq:nrqcd-S1gg}:
\begin{equation}
\begin{array}{l}
\displaystyle 
K_{P\bf{8}} =  v^2_{\mbox{eff}} = 0.15,\\
\\
K_{S\bf{8}} = 
K_{S\bf{1}}=\left[v^2_{\mbox{eff}}\right]^2 =0.0225.\\
\label{eq:K-choice}
\end{array}
\end{equation}
In order to estimate the uncertainties of the additional contributions to hadronic production  we vary the $K_{P\bf{8}}$ value from $0.1$ to $0.2$, and the 
$K_{S\bf{8,1}}$ value from $0.015$ to $0.03$.

It should be emphasized that the normalization values proposed in Eqs. \eqref{eq:nrqcd-P8g} to \eqref{eq:K-choice} cannot be regarded as reliable, and may drastically differ from values, which be will measured experimentally or predicted within a more rigorous approach. Nevertheless, we think that it is useful to demonstrate in this study, how the color octet contribution could influence the kinematic behavior of the $D$-wave $B_c$ meson yield at LHC experiments.

The calculation results  have been tested for Lorentz  invariance and gauge invariance. As it was already noted, the calculations were conducted within the method very close to ones applied to the study of $S$- and $P$-wave states. As in our previous researches the integration over phase space was carried out within the RAMBO algorithm~\cite{Kleiss:1985gy}.

While results of the latest calculations have been verified many times by other research groups~\cite{Chang:1994aw, Berezhnoy:1995au,Kolodziej:1995nv,Baranov:1997sg, Baranov:1997wv, Berezhnoy:1997uz, Chang:2004bh,Chang:2005wd,Chang:2006xka,Gao:2010zzc} we have tried to minimize the possibility of error in our new work.

\section{Estimations of relative yield }
For the numerical estimations of cross sections we involve the wave functions and masses listed in Table~\ref{tab:bc_input} (see also \Cref{tab:BcD} where the predictions for masses of $D$-wave excitations are presented). The parameters of radial wave functions for $1S$ and $1D$ states are taken from works~\cite{Eichten:2019gig}, \cite{Ebert:2011jc,Galkin:2020BcDwf}, and \cite{Berezhnoy:2019jjs,Martynenko:2021BcDwf}. Following most of the previous research on this topic, we choose the mass values of quarks in such a way as the mass of final $\bar b c$ quarkonium is correct. We are motivated by the fact, that relative yield of $2S$ excitations was described quite well within this choice.

\begin{table}[t]
\centering
 \caption{$B_c$-meson parameters involved in calculations.}
\label{tab:bc_input}
\begin{tabular}{|c|c|c|c|c|c|}
\hline
$B_c$-states & $m_b$ & $m_c$  & \parbox{3cm}{\vspace*{0.3cm} $|R(0)|^2$, $|R''(0)|^2$  \\ \cite{Eichten:2019gig} \vspace*{0.3cm}} & \parbox{3.5cm}{\vspace*{0.3cm} $|R(0)|^2_\textrm{eff}$, $|R''(0)|^2_\textrm{eff}$  \\ \cite{Ebert:2011jc,Galkin:2020BcDwf} \vspace*{0.3cm}} & \parbox{3.5cm}{\vspace*{0.3cm} $|R(0)|^2_\textrm{eff}$, $|R''(0)|^2$  \\ \cite{Berezhnoy:2019jjs,Martynenko:2021BcDwf}}

\\ \hline 
$1S$  & $~~4.80 \mbox{ GeV}~~$  & $~~1.50 \mbox{ GeV}~~$  &  $1.994 \mbox{ GeV}^3$ & $1.49 \mbox{ GeV}^3$ & $0.74 \mbox{ GeV}^3$ \\

$1D$ & $5.20\mbox{ GeV}$ &  $1.80\mbox{ GeV}$ &  $0.0986 \mbox{ GeV}^7$ & $0.116 \mbox{ GeV}^7$ & $0.055 \mbox{ GeV}^7$ \\
\hline
\end{tabular}
\end{table}

\begin{figure}[t]
\centering
\begin{minipage}{0.5\linewidth}
\centering
%\captionsetup{type=table,labelfont=bf,labelsep = period}
\includegraphics[width=\linewidth]{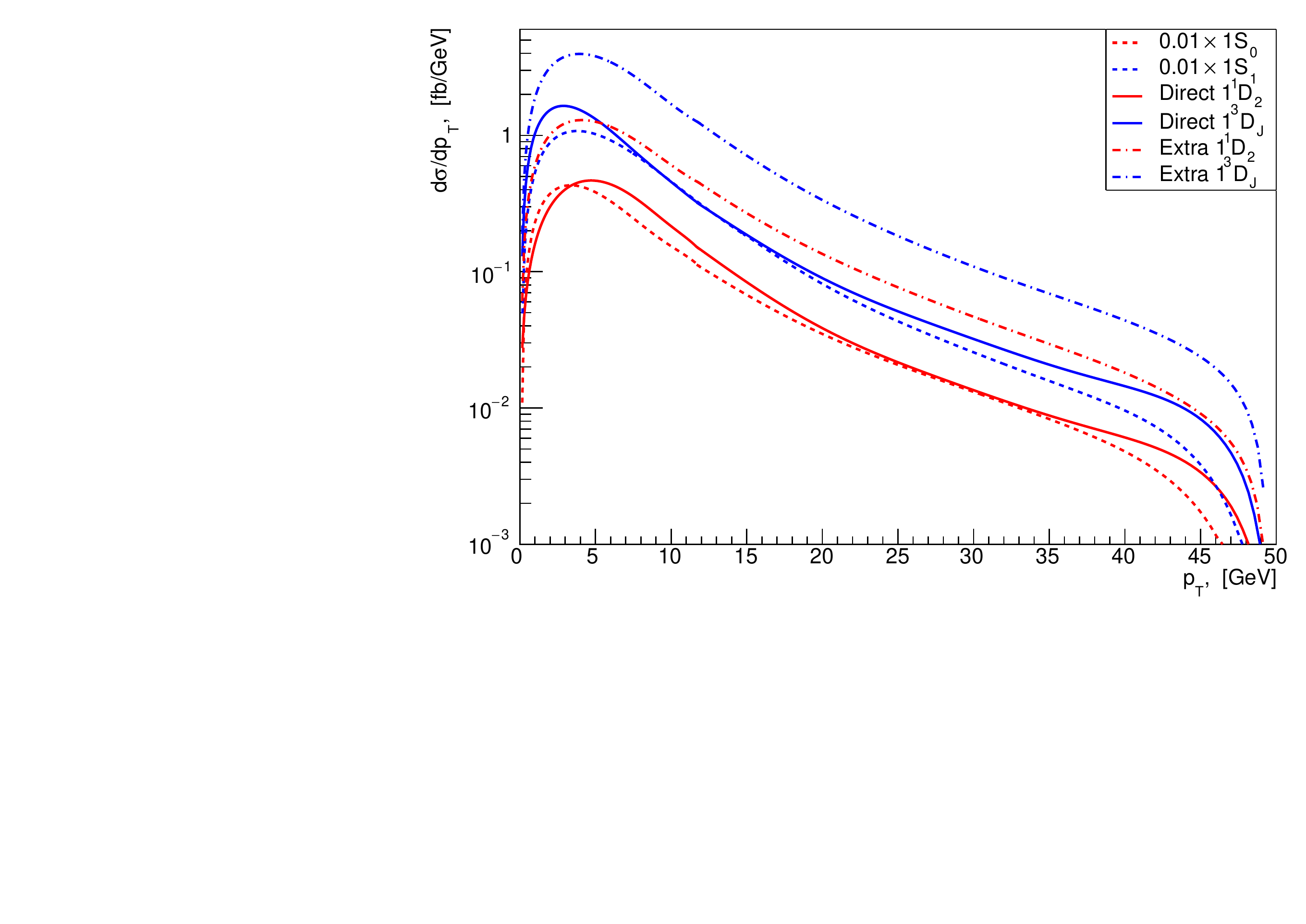}
\caption{$\sigma\left(gg\to B_c+X\right)$ dependence on transverse momentum at $\sqrt{s}_{gg}=100~\text{GeV}$. Solid lines: direct $D$-wave states; dashed lines: $S$-wave states scaled by $0.01$; dashed-dotted lines: extra $D$-wave states.
}
\label{fig:gg_pt}
\end{minipage}
\hfill
\begin{minipage}{0.49\linewidth}
\centering
%\captionsetup{type=table,labelfont=bf,labelsep = period}
\includegraphics[width=\linewidth]{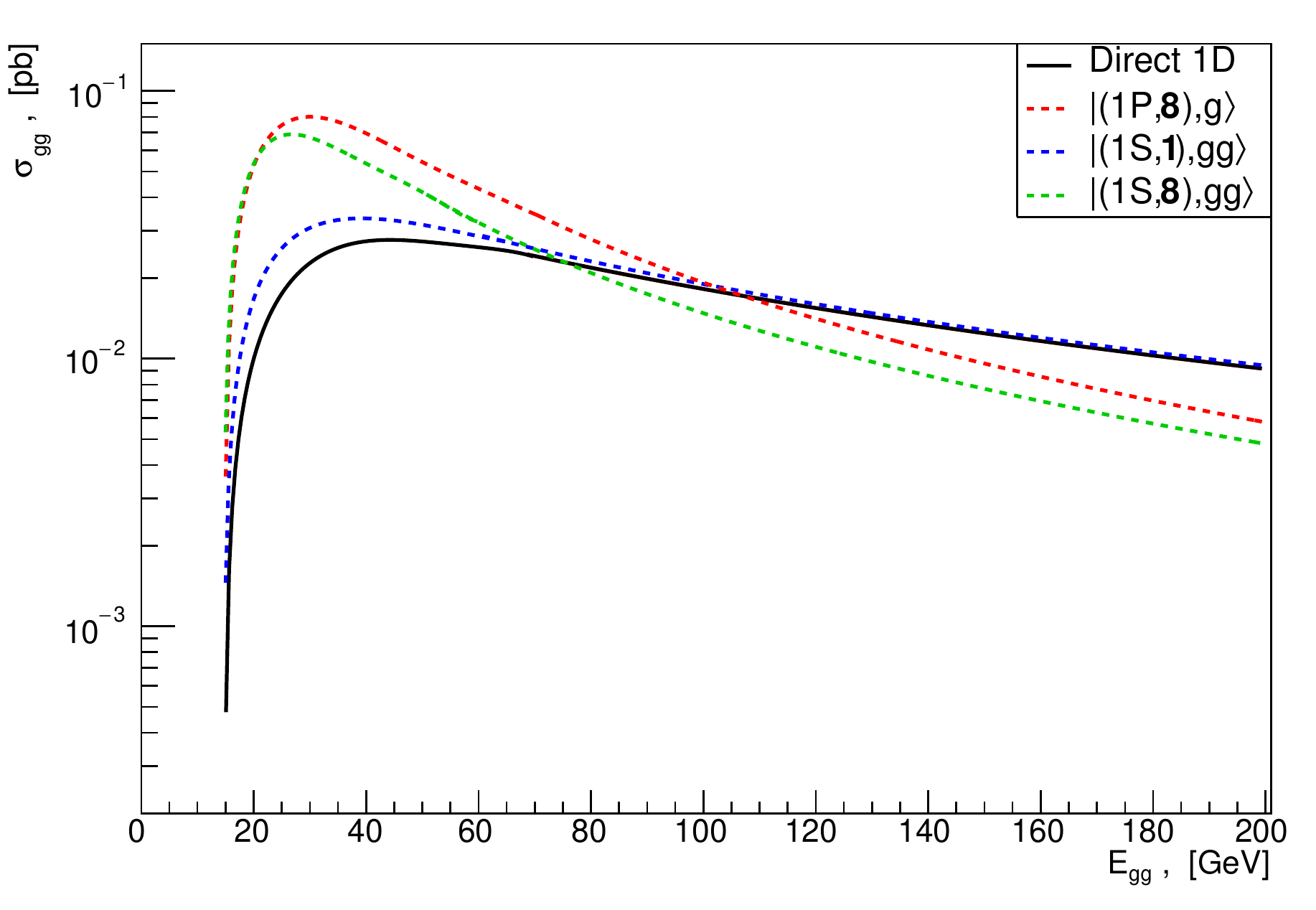}
\caption{$\sigma\left(gg\to B_c+X\right)$ dependence on gluon-gluon energy. Black line: direct $D$-wave production; red line:  $|(1P,{\boldsymbol 8})g\rangle$ contribution; blue and green lines: $|(1S,{\boldsymbol 1})gg\rangle$ and $|(1S,{\boldsymbol 8})gg\rangle$ contributions correspondingly.
}
\label{fig:gg_energy}
\end{minipage}
\end{figure}

Using the wave function parameters from  \cite{Eichten:2019gig} and choosing the quark masses as in \Cref{tab:bc_input} we predict, that the relative yield of $B_c(1D)$ with respect to the direct  $B_c(1S)$ in the gluonic fusion is about $0.5\div 1.3\ \%$, as seen from  \Cref{tab:gg2Bc}, where the cross section values for the gluonic production are presented at different gluonic energies. As  shown in Figure~\ref{fig:gg_pt} the distributions over transverse momentum for $D$-wave states are quite similar to ones for  $S$-wave states. It is worth mentioning, that the predicted ratio  of states with spin $S=1$ ($1 ^3D_1 $, $1 ^3D_2$,  $1 ^3D_3$)  v.s. states with spin $S=0$ ($1 ^1D_2$) is in approximate accordance with a simple spin counting rule (see ratios in Table~\ref{tab:LHC_yields}):

\begin{equation}
\frac{\sigma(1 ^3D_1 + 1 ^3D_2 + 1 ^3D_3)}{\sigma(1 ^1D_2 )}\sim  \frac{3+5+7}{5} = 3.
\end{equation}

This feature allows us to  use the prediction of quasipotential model~\cite{Ebert:2011jc,Galkin:2020BcDwf} where wave functions are  essentially different for  $1^3D_1$, $1^3D_2$, $1^3D_3$ and $1^1D_2$ states, as well as for  $1^3S_1$ and $1^1S_0$ states, even if contributions of different $D$ states are not estimated separately.  For this case we can approximately   estimate the cross section ratio averaging the wave function values according to spin counting rules (see Table~\ref{tab:bc_input}): 
\begin{equation}
 {|R_D''(0)|^2}_{\mbox{eff}}= \frac{3|R_{1^3D_1}''(0)|^2+5|R_{1^3D_2}''(0)|^2+7|R_{1^3D_3}''(0)|^2+5|R_{1^1D_2}''(0)|^2}{20},
 \label{eq:R_D_eff}
\end{equation}

\begin{equation}
 {|R_S(0)|^2}_{\mbox{eff}}= \frac{|R_{1^1S_0}''(0)|^2+3|R_{1^3S_1}(0)|^2}{4}.
 \label{eq:R_S_eff}
\end{equation}
Usage of \eqref{eq:R_D_eff} and \eqref{eq:R_S_eff} calculated within the approach  \cite{Ebert:2011jc,Galkin:2020BcDwf} leads to a little bit more optimistic values: the discussed relative yield is approximately $1.6$ times higher. 
Another model, based on the quasipotential approach~\cite{Berezhnoy:2019jjs,Martynenko:2021BcDwf}, predicts essentially lower values for wave functions at origin (see  Table~\ref{tab:bc_input} and Table~\ref{tab:bc_Galkin} of Appendix). However this difference almost disappears when estimating the relative yield. Using  the wave functions from ~\cite{Berezhnoy:2019jjs,Martynenko:2021BcDwf} increases the final value by  $1.5$ times comparing to \cite{Eichten:2019gig}.

\begin{table}[ht]
\centering
%\captionsetup{type=table}%,labelfont=bf,labelsep = period}
\vspace{2.5ex}
\caption{Gluonic cross sections at different energies; values are performed with $\alpha_S=0.1$ and wave functions from~\cite{Eichten:2019gig}.}
\begin{tabular}{|c|c|c|c|c|c|}
\hline
\multirow{2}{*}{
%\begin{tabular}[c]{@{}l@{}}
 $\sqrt{s}_{gg}$, GeV
%\end{tabular}
} & \multicolumn{5}{c|}{ $\sigma_{gg}$, pb} \\ \cline{2-6} 
    & $\ \ \ 1S\ \ \ $   &        $\ \ \ 1D\ \ \ $     & $|\bar b c( P , \boldsymbol{8} )g \rangle $  & $|\bar b c( S, \boldsymbol{8})gg \rangle$ & $|\bar b c(S,
\boldsymbol{1})gg \rangle$\\ \hline
20  &  1.97  &       0.009     &  0.051   &  0.053  & 0.016  \\ \hline
30  &  2.90  &       0.023     &  0.080   &  0.068  &  0.031  \\ \hline
50  &  2.64  &       0.028     &  0.055   &  0.042  &  0.032  \\ \hline
%60: 0.0259 +-0.0003
70  &  1.98  &       0.024     &  0.035   &  0.026  &  
%70: 0.0240 +- 0.0004
0.026  \\ \hline
%80: 0.0220 +- 0.0005
100 &  1.44  &       0.018     &  0.019   &  0.015  &  0.019  \\ 
%0.0184 +- 0.0006
\hline
150 &  0.904  &      0.012     &  0.010  &  0.007  &  0.013  \\ \hline 
%%0.0065
%200 GeV: 0.010+-0.001
\end{tabular}
\label{tab:gg2Bc}
\end{table}

As it is seen from \Cref{tab:gg2Bc} and Figs.~\ref{fig:gg_pt} and \ref{fig:gg_energy} the additional NRQCD contributions $|\bar b c( P , \boldsymbol{8} )g \rangle$,   $|\bar b c( S, \boldsymbol{8})gg \rangle$ and $|\bar b c(S,
\boldsymbol{1})gg \rangle$ extracted within naive velocity scaling rules (\ref{eq:nrqcd-P8g}, \ref{eq:nrqcd-S8gg},  \ref{eq:nrqcd-S1gg})  are able to crucially  change the production properties. While the $p_T$ distribution shapes are more or less the same, the energy dependencies for the  color octet contributions and for the color singlet contributions essentially differ: the color octet contributions decrease faster with energy, than the color singlet ones.  Moreover, seems, that the shape of energy dependence is mostly determined  by the color state of $\bar b c$-pair and practically does not depend on its orbital momentum. 

Concerning the numerical values of the additional NRQCD contributions one can conclude that each of such contributions is  of order of the direct color singlet production, as expected. This circumstance testifies to the self-consistency of our calculations. Since there are three such additional contributions, they crucially  increase the total cross section. 

To obtain the proton-proton cross sections the gluonic cross sections are convoluted with  CT14 PDFs~\cite{Dulat:2016rzo}:
\begin{equation}\label{sigma_pp}
\sigma_{pp} = \int\sigma_{gg}(\hat{s}_{gg},\mu)f_{g1}(x_1,\mu)f_{g2}(x_2,\mu)dx_1dx_2.
\end{equation}

To decrease uncertainties due to the scale choice and QCD corrections we present a relative yield of $1D$ states with respect to $1S$ states. The calculations are performed for  forward and central kinematic regions. 
The forward one is restricted by cuts $2<\eta<4.5,\ p_T<10~\text{GeV}$ and nearly corresponds to LHCb conditions, while the central one is restricted by cuts  $2<\eta<4.5,\ p_T<10~\text{GeV}$, $|\eta|<2.5,\ 10~\text{GeV}<p_T<50~\text{GeV}$ and approximately corresponds to CMS or ATLAS conditions. The proton-proton energy of collision is chosen equal to $13~\text{TeV}$.

When following the collinear approximation, one should always keep in mind the problem of transverse momentum of the initial gluons. Indeed, in some cases accounting the initial gluon transverse momenta crucially changes the production features (see, for example~\cite{Boer:2012bt} or  \cite{Likhoded:2016zmk}).  However, we believe that in our case  the dependence on  the initial gluonic transverse momenta is more or less eliminated in the ratio $\sigma(B_c(D))/\sigma(B_c(S)$.

\begin{figure}[t]
\centering
\begin{minipage}{0.49\linewidth}
\centering
\includegraphics[width=\linewidth]{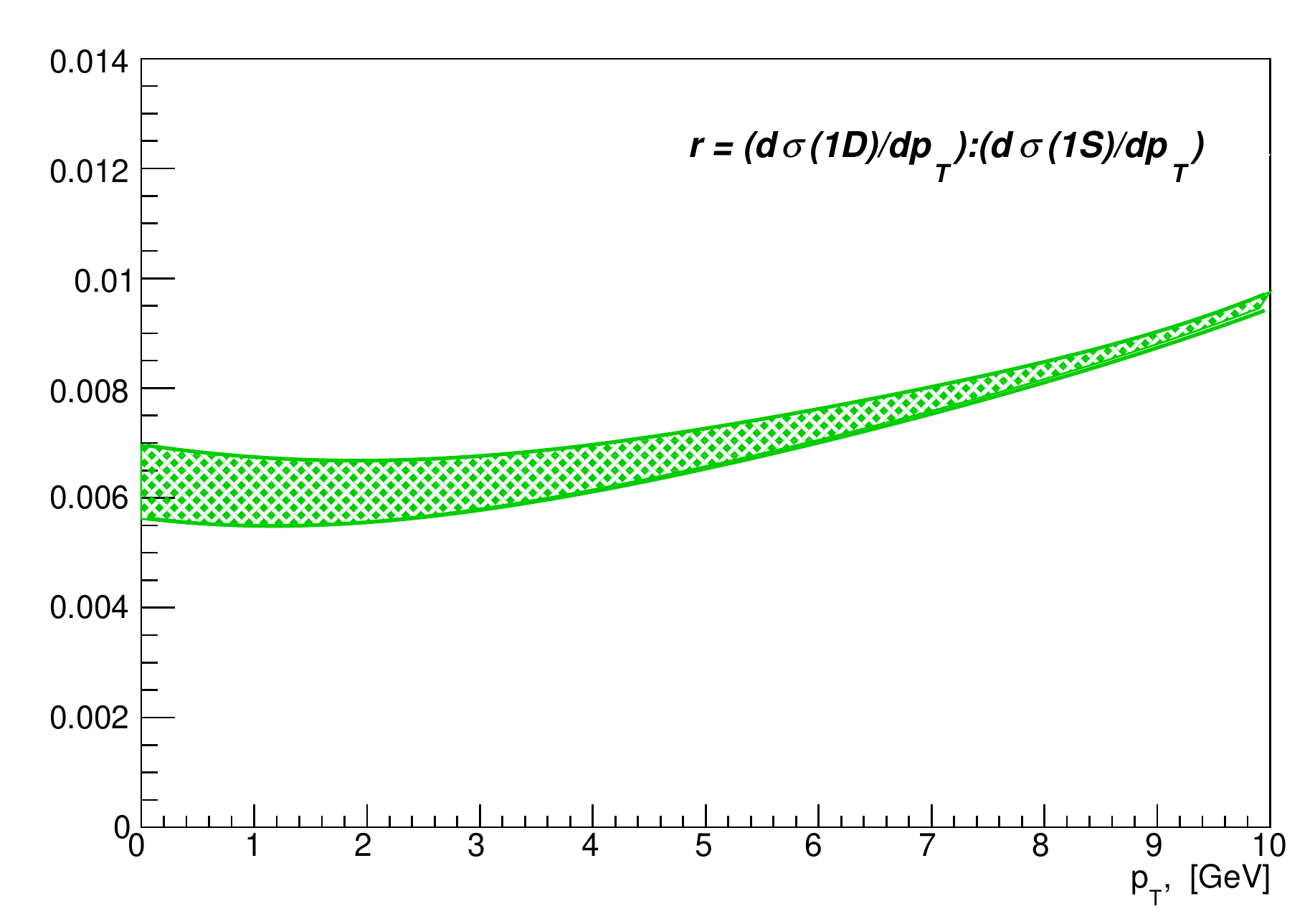}
\caption{ $r$ dependence on $p_T$ at different scales for forward kinematics: $2<\eta<4.5,\ p_T<10~\text{GeV}$.}
\label{fig:pt_ratio_LHCb}
\end{minipage}
\hfill
\begin{minipage}{0.49\linewidth}
\centering
\includegraphics[width=\linewidth]{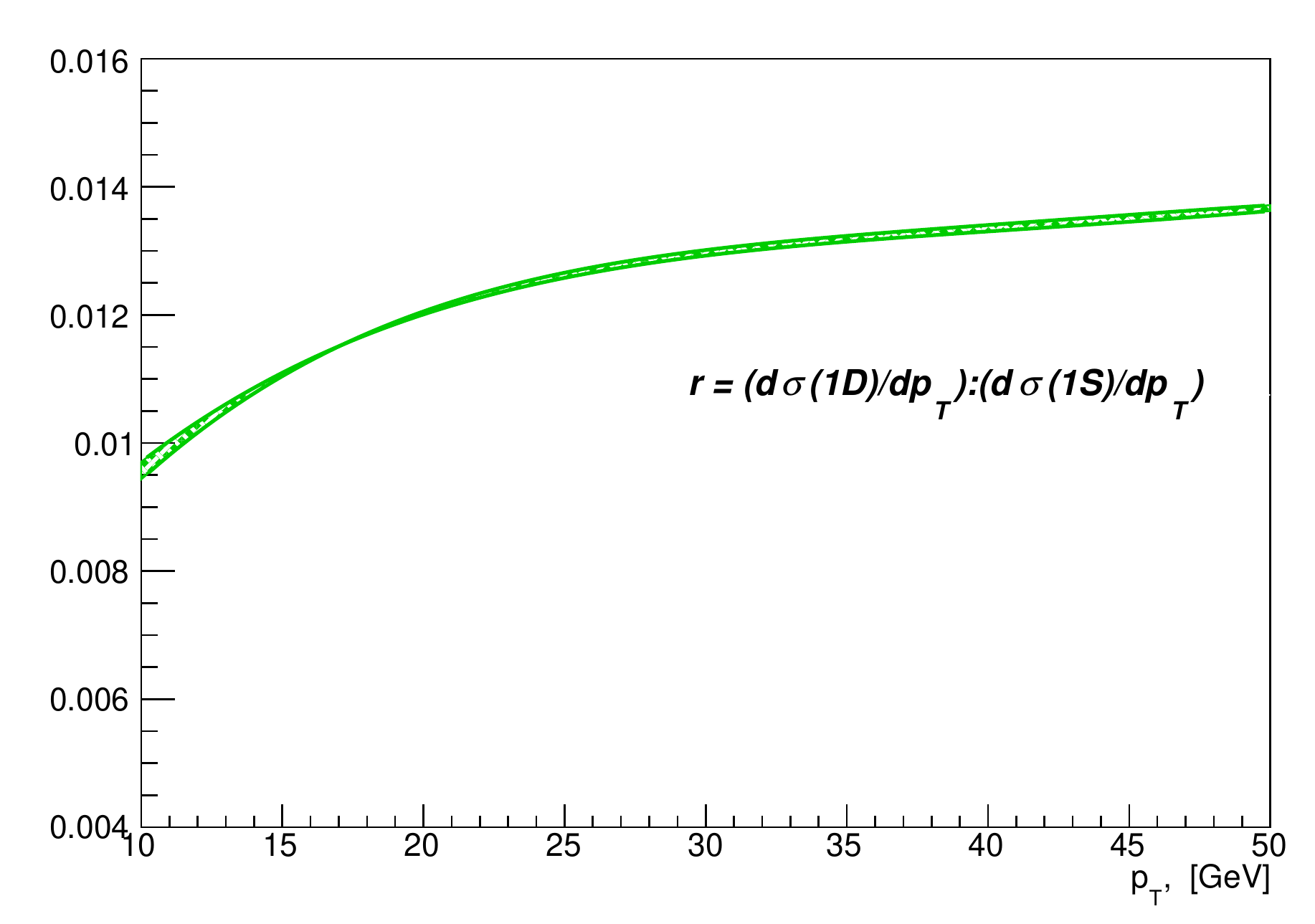}
\caption{ $r$ dependence on $p_T$ at different scales for central kinematics: $|\eta|<2.5,\ 10~\text{GeV}<p_T<50~\text{GeV}$.}
\label{fig:pt_ratio_CMS}
\end{minipage}
\end{figure}

The systematic uncertainty of the calculations is estimated with variation of the scale in the range $E_T/2<\mu <2E_T$. It is well seen in  \Cref{fig:pt_ratio_LHCb,fig:pt_ratio_CMS} that the relative yields are hardly dependent on scale variation.  As it is seen in~\Cref{tab:LHC_yields} within the applied model (the color singlet production in the gluon fusion subprocess) the relative yield value depends on kinematics: for the central region it is approximately twice as large. The use of the wave functions set~\cite{Ebert:2011jc,Galkin:2020BcDwf} or  \cite{Berezhnoy:2019jjs,Martynenko:2021BcDwf} increases the predicted yield of $D$-wave states to $1\div 1.8$ \%.

\begin{figure}[t]
\centering
\begin{minipage}{0.49\linewidth}
\centering
\includegraphics[width=\linewidth]{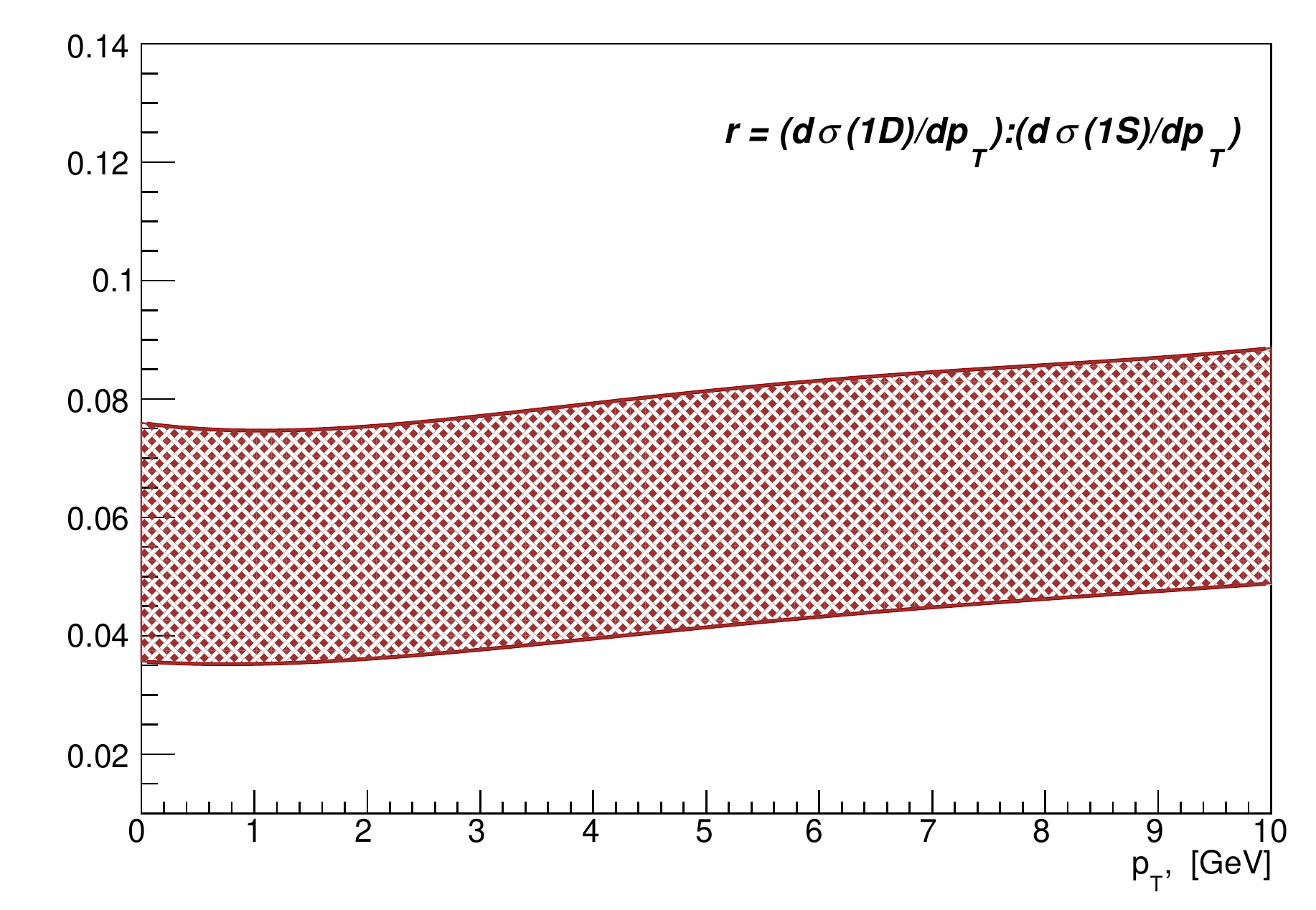}
\caption{ $r$ dependence on $p_T$ at different scales for forward kinematics: $2<\eta<4.5,\ p_T<10~\text{GeV}$. The contributions of   $|\bar b c( P , \boldsymbol{8} )g \rangle$,   $|\bar b c( S, \boldsymbol{8})gg \rangle$ and $|\bar b c(S,
\boldsymbol{1})gg \rangle$ states are included.}
\label{fig:pt_ratio_LHCb_oct}
\end{minipage}
\hfill
\begin{minipage}{0.49\linewidth}
\centering
\includegraphics[width=\linewidth]{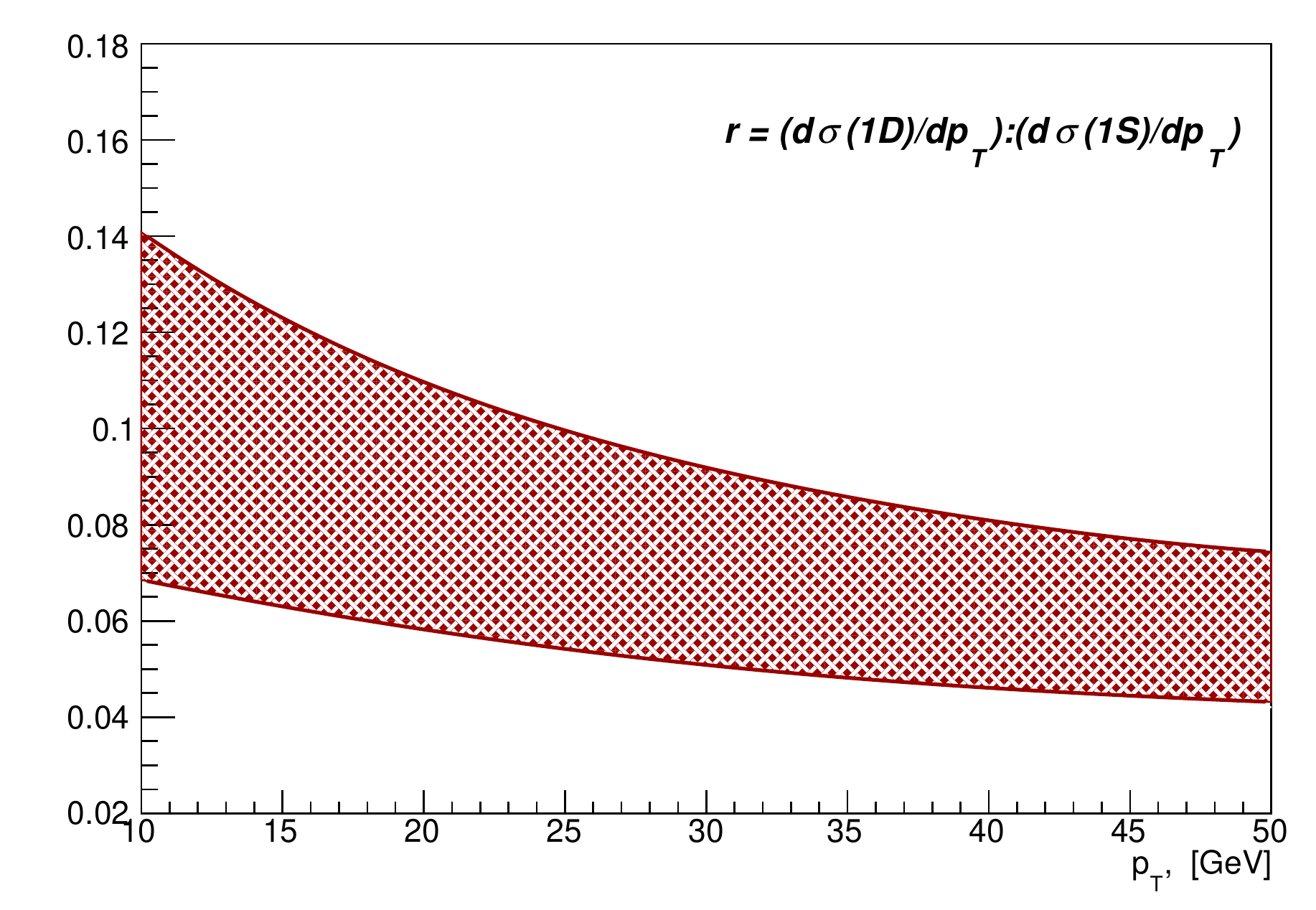}
\caption{ $r$ dependence on $p_T$ at different scales for central kinematics: $|\eta|<2.5,\ 10~\text{GeV}<p_T<50~\text{GeV}$. The contributions of   $|\bar b c( P , \boldsymbol{8} )g \rangle$,   $|\bar b c( S, \boldsymbol{8})gg \rangle$ and $|\bar b c(S,
\boldsymbol{1})gg \rangle$ states are included.}
\label{fig:pt_ratio_CMS_oct}
\end{minipage}
\end{figure}

Accounting naively estimated contributions of   $|\bar b c( P , \boldsymbol{8} )g \rangle$,   $|\bar b c( S, \boldsymbol{8})gg \rangle$ and $|\bar b c(S, \boldsymbol{1})gg \rangle$ can increase the relative yield  of $D$-wave states by an order of magnitude, as it is shown in Figs.~\ref{fig:pt_ratio_LHCb_oct} and \ref{fig:pt_ratio_CMS_oct}.  Despite the fact that each additional contribution is comparable to the main one, their number provides such an increase in yield. As noted in the previous section we estimate the uncertainties for these contributions varying  $K_{P\bf{8}}$ from $0.1$ to $0.2$, and $K_{S\bf{8,1}}$ from $0.015$ to $0.03$ [see (\ref{eq:nrqcd-P8g}, \ref{eq:nrqcd-S8gg}, \ref{eq:nrqcd-S1gg})].

We  emphasize ones again, that the  naive  normalization used in this research  cannot be regarded as reliable, and may be drastically far from values, which will be measured experimentally or predicted within a more rigorous approach.

\begin{table}[t]
    \centering
    \caption{Relative yields for $D$-wave $B_c$ mesons for forward and central kinematic regions at LHC; the wave functions set \cite{Eichten:2019gig} is applied.}
    \begin{tabular}{|c|c|c|c|}
    \hline
 kinematic region    &  $\sigma\left(1^3S_1\right)/\sigma\left(1^3S_0\right)$ & $\sigma\left(1^3D_j\right)/\sigma\left(1^1D_2\right)$ &$\sigma\left(1D\right)/\sigma\left(1S\right)$, \% \\\hline
 $2<\eta<4.5,\ p_T<10~\text{GeV}$ & 2.4 & 3.0 &  0.6 $\div$ 0.7 \\   
 $|\eta|<2.5,\ 10~\text{GeV}<p_T<50~\text{GeV}$     
 & 2.4 &  2.3 &  1.0 $\div$ 1.1\\  \hline
    \end{tabular}
    \label{tab:LHC_yields}
\end{table}

\section{Conclusions}
 The $B_c(2S)$ excitations have been observed at LHC in the $B_c \pi^+\pi^-$ spectrum~\cite{Sirunyan:2019osb, Sirunyan:2020mzn,CMS:2020wnn, Aaij:2019ldo}, and this result stimulated us to estimate possibilities to search for $B_c(D)$ excitations in the same spectrum.  At very large statistics it would be possible to distinguish two peaks in the $B_c\pi^+\pi^-$ mass spectrum: one peak  near $7000$~GeV formed by $1 ^1D_2$ state  and another  one  near $6930$~GeV  formed by $1^1D_1$, $1^1D_2$ and $1^1D_3$ states decaying to  $B_c^* \pi^+\pi^-$ with further radiative decay $B_c^* \xrightarrow{\gamma} B_c$. Also the $D$-wave $B_c$ excitations could be found in cascade radiative decays $B_c(1D)\xrightarrow{\gamma} B_c(1P) \xrightarrow{\gamma} B_c(1S)$.
 
Taking into account the main color singlet contribution, we estimate $B_c(D)$ states yield in the hadronic production as $0.6\div 1.8$\% with respect to the direct production of $1S$ states for the chosen mass values. To convert  this ratio into the  more representative ratio of $D$-wave states yield to the yield of all $B_c$ mesons, one should  divide it by a factor of about 1.5, that leads to the values $0.4\div 1.1$\%. Our estimations of the relative yield of $D$-wave $B_c$ states in the hadronic production   do not contradict the analogous estimations within the fragmentation approach~\cite{Cheung:1995ir}. 

Accounting  contributions of  $|\bar b c( P, \boldsymbol{8})g \rangle$, $|\bar b c( S, \boldsymbol{8})gg \rangle$ and $|\bar b c(S, \boldsymbol{1})gg \rangle$ states extracted using the naive velocity scaling rules  increases the relative yield  of $D$-wave states by an order of magnitude. Therefore the  significant experimental excess  of the relative yield  of $D$-wave mesons over the  value $0.4\div 1.1$\% will indicate an essential contribution of the color octet states to the production.
 
We have to conclude that an observation of the discussed states at LHC is a quite challenging experimental task due to the small relative yield.

The authors would like to thank V. Galkin and A. Martynenko for help and useful discussions. The work was supported by RFBR (grant No. 20-02-00154 A). The work of I.~Belov was supported by ``Basis'' Foundation (grant No. 20-2-2-2-1). 

\clearpage
\appendix*
\section{Wave functions and spectroscopy of the $D$-wave $B_c$ states}

In this Appendix we present the  masses of $D$-wave states of $B_c$ meson predicted within different models~\cite{Eichten:1994gt, Gershtein:1994jw, Zeng:1994vj, Fulcher:1998ka, Ebert:2002pp, Godfrey:2004ya, Monteiro:2016ijw, Soni:2017wvy, Li:2019tbn}. 
As well we present the wave function parameters obtained within the quasipotential  approaches~\cite{Ebert:2011jc,Galkin:2020BcDwf} and~\cite{Berezhnoy:2019jjs,Martynenko:2021BcDwf}.  
\begin{table}[ht]
\caption{Predictions for masses of $D$-wave $B_c$ meson states in MeV.}
\label{tab:BcD}
\begin{tabular}{||l|c|c|c|c|c|c|c|c|c||}
\hline
State & EQ~\cite{Eichten:1994gt} & GKLT~\cite{Gershtein:1994jw} & ZVR~\cite{Zeng:1994vj} & FUI~\cite{Fulcher:1998ka} & EFG~\cite{Ebert:2002pp} & GI~\cite{Godfrey:2004ya}  & MBV~\cite{Monteiro:2016ijw} & SJSCP~\cite{Soni:2017wvy} & LLLGZ~\cite{Li:2019tbn}\\
 \hline
  $1^1D_2$ & 7009 & \ldots & 7020 & 7023 & \ldots & \ldots & \ldots &  6994 & \ldots \\ 
$1^3D_1$ & 7012 & 7008 & 7010 & 7024 & 7072 & 7028 & 6973 & 6998 & 7020\\
 $1^3D_2$ & 7012 & \ldots & 7030 & 7025 & \ldots & \ldots & \ldots & 6997 & \ldots \\
 $1^3D_3$ & 7005 & 7007 & 7040 & 7022 & 7081 & 7045 & 7004 & 6990 & 7030\\
$1~D_2'$ & \ldots & 7016 & \ldots & \ldots & 7079 & 7036 & 7003 & \ldots & 7032\\
 $1~D_2$ & \ldots & 7001 & \ldots & \ldots & 7077 & 7041 & 6974 & \ldots & 7024\\ \hline
 \end{tabular}
\end{table}

\begin{table}[ht]
\centering
 \caption{$B_c$ meson wave functions within the quasipotential models \cite{Ebert:2011jc,Galkin:2020BcDwf} and \cite{Berezhnoy:2019jjs,Martynenko:2021BcDwf}.}
\label{tab:bc_Galkin}
\begin{tabular}{|c|c|c|}
\hline
$B_c$-state & $|R(0)|^2$, $|R''(0)|^2$~\cite{Ebert:2011jc,Galkin:2020BcDwf} & $|R(0)|^2$, $|R''(0)|^2$~\cite{Berezhnoy:2019jjs,Martynenko:2021BcDwf}                \\ \hline
$1^1S_0$    & $2.68 \mbox{ GeV}^3$                                                          & $0.97 \mbox{ GeV}^3$                    \\ %\hline
$1^3S_1$    & $1.09 \mbox{ GeV}^3$                                                          & $0.66  \mbox{ GeV}^3$                   \\ \hline
$1^1D_2$    & $0.078 \mbox{ GeV}^7$                                                         & \multirow{4}{*}{$0.055  \mbox{ GeV}^3$} \\ %\cline{1-2}
$1^3D_1$    & $0.314 \mbox{ GeV}^7$                                                         &                                         \\ %\cline{1-2}
$1^3D_2$    & $0.098 \mbox{ GeV}^7$                                                         &                                         \\ %\cline{1-2}
$1^3D_3$    & $0.061 \mbox{ GeV}^7$                                                         &                                         \\ \hline
\end{tabular}\end{table}

\clearpage
\bibliography{Bc_excitations}

\end{document}